\documentclass[a4paper,11pt]{article}
\pdfoutput=1 % if your are submitting a pdflatex (i.e. if you have
\usepackage{jcappub} % for details on the use of the package, please
\usepackage[T1]{fontenc} % if needed

\usepackage{amssymb, amsmath, epsfig, natbib}

\author[a]{Martin White}
\author[b]{Nikhil Padmanabhan}

\affiliation[a]{Department of Physics, University of California,
Berkeley, CA 94720}
\affiliation[b]{Department of Physics, Yale University, New Haven, CT 06520}

\emailAdd{mwhite@berkeley.edu}
\emailAdd{Nikhil.Padmanabhan@yale.edu}

\title{Including parameter dependence in the data and covariance for
cosmological inference}


\abstract{The final step of most large-scale structure analyses involves the
comparison of power spectra or correlation functions to theoretical models.
It is clear that the theoretical models have parameter dependence, but
frequently the measurements and the covariance matrix depend upon some of the
parameters as well.  We show that a very simple interpolation scheme from an
unstructured mesh allows for an efficient way to include this parameter
dependence self-consistently in the analysis at modest computational expense.
We describe two schemes for covariance matrices.
The scheme which uses the geometric structure of such matrices performs
roughly twice as well as the simplest scheme, though both perform very well.
}

\begin{document}
\maketitle
\flushbottom

\section{Introduction}

The study of large-scale structure - using the cosmic microwave background,
galaxy and cluster surveys, weak lensing, 21cm-background fluctuations and
other probes - promises to teach us a wealth of information about our Universe
and theories of fundamental physics and for this reason has been the focus of
intense community effort for many decades.
The final stage in any cosmological data analysis, from which the key
constraints and insights are derived, is the comparison of a summary
statistic (or set of statistics) with the predictions of a theoretical
model, and it is this stage which is the focus of this paper.
Within cosmology the most common approach is to construct a likelihood
function and derive limits on a set of cosmological parameters (or a class
of theories) given the data.  Frequently this is done with a Markov Chain
Monte Carlo algorithm \cite{Gil96,Gil99}
or nested sampling scheme \cite{CosmoNest}.
What is desired then is the ability to compute the likelihood function of
the data, given the model, as a function of the theoretical parameters.

Within the large-scale structure community the most common assumption about
the likelihood function is that it is Gaussian.
This is inspired by the central limit theorem under the assumption that there
are many modes contributing to each measurement.
The approximation becomes more reliable as the data become better constraining,
though its reliability can depend on the summary statistic being used to
compare data with theory.\footnote{For example, it is likely to be a worse
approximation for the power spectrum than for the correlation function,
since the power spectrum in a bin is the sum of positive definite quantities.}
For this paper, we shall assume that the Gaussian likelihood approximation is
sufficient; we refer the reader to the existing literature 
on the scope and validity of this approximation for different probes.

We denote the parameters upon which our theory depends with $\vec{p}$ and
write the data in vector form, $d_i$ and expected value and covariance 
of the observations as $\mu_i(\vec{p})$ and $C_{ij}(\vec{p})$.
Within the Gaussian approximation, the likelihood is 
\begin{equation}
  \mathcal{L}(\vec{p}) \propto \left| C(\vec{p})\right|^{-1/2}
  \exp\left[-\frac{1}{2}\left(d_i(\vec{p})-\mu_i(\vec{p})\right)^T
  C(\vec{p})^{-1}_{ij}\left(d_j(\vec{p})-\mu_j(\vec{p})\right)\right].
\end{equation}
We emphasize that dependence on $\vec{p}$ exists in three places -
(i) the prediction $\mu_i(\vec{p})$,
(ii) the data vector $d_i(\vec{p})$ and
(iii) the covariance matrix $C_{ij}(\vec{p})$.
Any {\it self-consistent} evaluation of the likelihood function needs
to address all three dependences.
The first of these is what might be described as the ``modeling''
or ``prediction'' phase and can be addressed with approaches from
direct analytic calculation to detailed numerical simulations.
We will assume that this dependence is addressed at the level of
precision required for current and future surveys. Our purpose in this
paper is to argue that the latter two dependences, often ignored in
analyses, are also readily addressable.

Motivated by studies of weak gravitational lensing, baryon acoustic
oscillations (BAO) and redshift space distortions (RSD), we are interested
in measuring the 2-point function of shear, galaxies, quasars or Lyman $\alpha$
absorption from a large (redshift) survey.
To simplify the exposition we shall specialize to the case of a galaxy
2-point function, though most of our statements are intended to hold more
generally.
The summary statistic is thus the galaxy power spectrum or correlation
function, or perhaps an integral of these (e.g.~\cite{Xu10}).
We have indicated a dependence of $d_i$ on $\vec{p}$ since, given the
observed positions of galaxies and the selection function, computation of
the correlation function or power spectrum requires the assumption of a
fiducial cosmology and this should ideally be consistent with the cosmological
parameters being tested.  Further model dependence is introduced if `weights'
are used in the calculation or if density field reconstruction \cite{Eis07}
is used to sharpen the acoustic features in the clustering signal.

In what follows we shall assume that $\mu_i(\vec{p})$ is in hand.
This is the `modeling' phase of the problem, and is the subject of a
large literature in itself.
Recent comparisons of models focused on BAO and RSD can be found in
\cite{Bla11,Bla12,And14,Whi15,ZelRecon}.
The computation of $C_{ij}(\vec{p})$ is more complex.
While the expression for Cov$[P(\vec{k}'),P(\vec{k}'')]$ or
Cov$[\xi(\vec{r}'),\xi(\vec{r}'')]$ within linear perturbation theory
(assuming Gaussian statistics for the density fluctuations) is
straightforward, deviations from linear theory can be important on the
scales of interest for current and future surveys and properly handling
the mask associated with a complex observing geometry is non-trivial.
The situation becomes even more complex if we use a non-linear filter such
as density field reconstruction (to sharpen the BAO peak) before computing
the 2-point function.  
These complexities have led to analysts generating their covariance matrices
through Monte Carlo simulation of a large number of mock catalogs.
While ideally the covariance matrix is generated for each and every
parameter set of interest, this is computationally expensive.  Even the
more limited procedure of iteratively regenerating the covariance matrix
for the best fit cosmology is not usually attempted.
As we prepare for the next generation of surveys, it behooves us to rethink
these steps.
In what follows we shall advocate and investigate a low order interpolation
scheme that provides a simple, approximate, route to including the
parameter variations in $d_i$ and $C_{ij}$ with moderate computational cost.

Essentially our proposal is to generate an interpolator (or emulator, or
response model) for the data and covariance matrix based on values
pre-computed at certain points in the parameter space.
This interpolator then allows the rapid computation of the likelihood for
any cosmology, self-consistently including changes to the fiducial cosmology,
bias and growth factor.
We implicitly assume that for current and future surveys the range of
parameters being searched are sufficiently small that a low order
interpolation is sufficient, and we investigate here a simple linear
interpolation from an unstructured mesh.
This is not the only choice, but it does allow for easy updating and refining
of the mesh as the data improve and the attention is focussed on smaller
regions of parameter space.
Recently \cite{Mor13} presented an alternative procedure, aimed at the weak
lensing case.  This builds upon work presented in \cite{CoyoteII} which
developed an interpolation for cosmological power spectra derived from
N-body simulations.
Some of the techniques introduced or refined in those papers could be of use
in our context, but we have found that the simple situation described below
performs quite well in this context so the more sophisticated approach may
not be needed.

The outline of the paper is as follows.
In the next section we describe the basics of our interpolation scheme.
We describe the simple case of interpolation of the data vector,
e.g.~$\xi_\ell(r_i)$ or $P(k_i)$, in section \ref{sec:interp_data}.
Section \ref{sec:interp_cov} describes how one can interpolate the
covariance matrix, or its inverse the precision matrix.  This situation
is more complex, so we divide this section into several subsections and
relegate some technical details to appendices.
We finish with a summary of our main results and directions for future
work in section \ref{sec:conclusions}.

\section{Interpolation scheme}
\label{sec:interp}

Below we shall discuss how to compute the data and covariance at particular
points in the parameter space, but first we introduce some notation which
will be used in each case.
To begin we distinguish between `fast parameters', which may enter the data,
model or covariance matrix but do not require any difficult recomputations
and `slow' parameters which need significant calculation.
We need only interpolate in the `slow' parameters, and henceforth we shall
restrict ourselves to this set.
We shall assume there are $n_p$ such parameters.
We shall also assume that we are interpolating a single object
(e.g.~the total covariance matrix rather than the sample variance and
shot-noise pieces of it) though the generalization to multiple objects
is trivial and splitting the data or covariance matrix may be beneficial.

There are numerous methods for performing interpolation
(schemes which are particularly relevant to our situation are kernel
interpolation and tensor B-splines)
and we do not need to use the same method for each component of the problem.
However doing so simplifies the discussion.
Perhaps the simplest technique is multilinear interpolation (from an
unstructured grid) and we shall use this method below as an illustrative
example.  The reasons are twofold.  First, we expect changes to our data
and covariance to be small and smooth, so a low-dimensional method should
be adequate and very easy to code.
Second, by using an unstructured mesh it is easy to later add additional
interpolation points if they become available.
Adding additional points close to the regions of high likelihood can be
more valuable than more complex interpolation schemes.
Of course, many of the points we discuss below can be adapted to more complex
interpolation schemes if necessary.

To interpolate from an unstructured mesh we need our functions evaluated for
at least $n_p+1$ points.
To begin let us imagine that we have the data, model or covariance evaluated
at $n_p+1$ points in the parameter space, chosen such that the simplex they
define encloses the entire high-likelihood region.\footnote{We note that this
is very unlikely to be the best distribution, but serves to introduce the
idea.}  We can trivially interpolate to any interior point using
(homogeneous) M\"{o}bius barycentric coordinates, also known as areal
coordinates \citep[e.g.][]{Cox69}.
Given a point, $\vec{p}$, interior to the simplex defined by $\{\vec{p}_i\}$,
solve for the areal coordinates $x_i$ which satisfy
$\vec{p}=\sum_i x_i \vec{p}_i$ and $\sum_i x_i=1$
(this is a simple linear transformation).
If the point is interior to the simplex then $0\le x_i\le 1$ for all $i$,
and this can be used as a test for an interior point.
Then the interpolated function, $F$, is simply $\sum_i x_i F_i$, where the
$F_i$ are the data, model or covariance evaluated at parameters $\vec{p}_i$.
This technique is frequently referred to as linear Lagrange interpolation.

\begin{figure}
\begin{center}
\resizebox{\columnwidth}{!}{\includegraphics{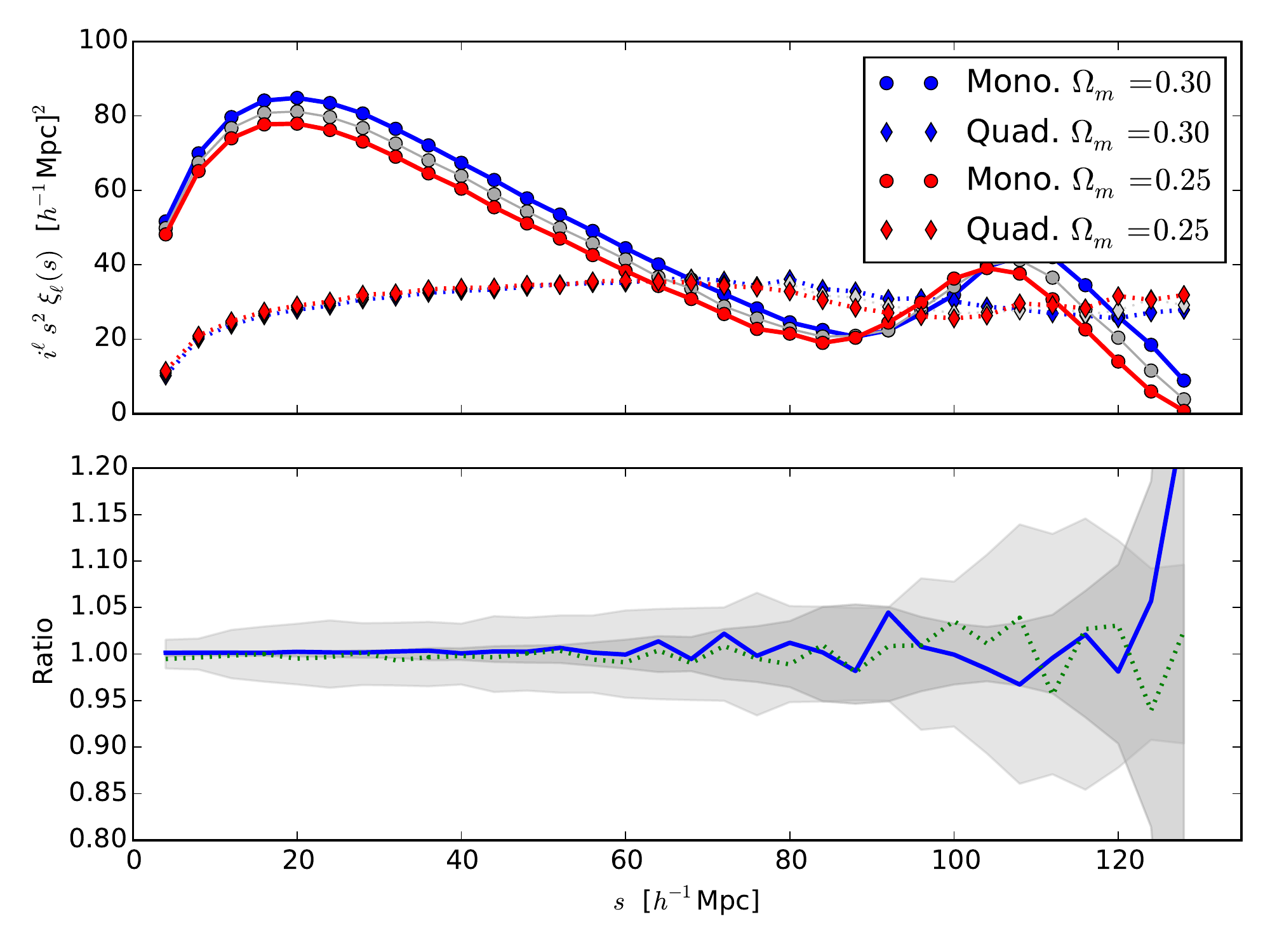}}
\caption{How the correlation function changes with distance scale.
In the upper panel the lines show the average redshift-space monopole
and quadrupole moments of the correlation function of halos from two N-body
simulations assuming a distance redshift relation appropriate to
$\Lambda$CDM with $\Omega_m=0.25$ (red), 0.275 (light grey) and 0.30 (blue).
The distance to $z=1$ varies by 3 per cent for these cosmologies.
The halos are taken from the $z=1$ outputs of two simulations with the same
cosmology but different initial conditions and observed from each of the 8
corners as if along a lightcone.  The halos have masses
$5\times 10^{12}<M<10^{13}\,h^{-1}M_\odot$ and form 16 realizations of
an octant of the sky cut to $0.8<z<1.2$.
Note that the correlation function assuming $\Omega_m=0.275$ can be
described to better than 10 per cent by the linear interpolation of the 0.25
and 0.30 results except where it crosses zero.
The lower panel shows the ratio of the linear interpolation to the 0.275
result.  The light and dark grey bands show the $1\,\sigma$ error on the
mean (from the scatter of the 16 samples) of the $\Omega_m=0.275$ value
for the quadrupole and monopole respectively.  The interpolation error is
below this for all scales even for such a large interpolation.}
\end{center}
\label{fig:xi-interp}
\end{figure}

A better scheme would add at least one additional point, close to a prior
guess of the maximum likelihood parameters.  Adding further points in this
region is desirable, with the number limited primarily by computational
resources.
There are three criteria which could determine where to place additional
points:
(1) place points where the likelihood is high, or
(2) in regions where the linear interpolation performs the worst, or
(3) near the edges of the parameter space (if the simplices so defined prove
advantageous).
The best distribution of points is likely to be dependent upon the specifics
of the problem.
In the presence of these additional points the interpolation begins by first
finding the $n_p+1$ points closest to the point, $\vec{p}$, of interest whose
simplex\footnote{This is often called the `enclosing Delaunay simplex', and
it can be shown that the Delaunay triangulation gives rise to the piecewise
linear approximation with the smallest maximum error at each point over
functions with bounded curvature \cite{Raj91}.
Since finding the bounding simplex is such a common problem
(e.g.~in collision detection and robot motion planning or computer graphics)
there is a large literature on the subject and many well developed packages
that can be used.  We have used the routines in
{\tt scipy.interpolate} and {\tt scipy.spatial}.}
encloses $\vec{p}$.  Using these $n_p+1$ points, find the areal coordinates,
$x_i$ and compute $\sum_i x_iF_i$ as before.
We note in passing that parameter redefinitions may improve the performance of
any interpolation scheme and that if we have a metric on the parameter space
we can define `near' and `far' in terms of an approximation to a prior
likelihood, but we shall not pursue this route here.
We also note that while we have pursued interpolation from an unstructured
mesh, many of the techniques we describe below can be useful in other
interpolation schemes.

\section{Interpolating the data vector}
\label{sec:interp_data}

Typically one treats the data vector as fixed when doing cosmological
inference, but in principle it can depend upon the cosmological model
if that model includes changes in the distance-redshift relation or
if complex processes such as density field reconstruction are applied
to the data prior to computing the power spectrum or correlation function.
For the range of cosmologies usually considered the changes in the
distance-redshift relation are relatively small, and smooth.  A similar
smoothness applies for density field reconstruction.
Thus we expect any changes in our 2-point function to be smooth functions
of the input parameters and hope that they can be linearly interpolated
with little error.

To illustrate this point we take the positions halos from two large N-body
simulations, viewed from each corner of the boxes as if they filled an octant
on the sky.  Sixteen samples of halos were constructed, each in a shell
$0.8<z<1.2$.
Holding the angular and redshift coordinates appropriate to the `true'
cosmology (as assumed in running the simulation) we convert to distances in
three other cosmologies with different distance-redshift relations and then
compute the halo auto-correlation function.
For simplicity we take the distance-redshift relation to be that of
$\Lambda$CDM cosmologies with $\Omega_m=0.25$, 0.275 and 0.3 though this is
just for illustration and when analyzing data it may make more sense to use
a scheme such as outlined in ref.~\cite{Zhu15}.
The results are shown in figure \ref{fig:xi-interp} where we see that the
changes are smooth functions of the input parameters and linear interpolation
{}from the endpoints results in errors in the interior smaller than the
observational error.  The variation in currently allowed cosmologies
is smaller than this, and thus the interpolation should perform even better
when used in cosmological inference, especially if one of the places where
the data are evaluated is very close to the best-fitting cosmology.
For current and future surveys there is little cost to computing the
correlation function or power spectrum of the data for a number of different
fiducial cosmologies, so the interpolation can be made almost arbitrarily
precise.  For this reason we shall not investigate this in detail, and turn
instead to the more complex step of interpolating the covariance matrix.
This is frequently the more computationally demanding calculation and is
also the least straightforward conceptually.

\section{Interpolating covariance}
\label{sec:interp_cov}

In large-scale structure analyses it is often the case that the covariance
matrix is fixed throughout the analysis, usually at a fiducial cosmology
where it was determined either through analytic means or, most often,
Monte Carlo simulation.
Computation of the covariance matrix from Monte Carlo simulation is often
the most computationally demanding step in the analysis.
Using the wrong covariance matrix won't lead to a bias on average, but it
may alter the confidence levels (see e.g.~ref.~\cite{Tay13} for a recent
study and see \cite{Mor13} for a discussion of computing parameter dependent
covariance matrices in the weak lensing case).
In the limit that the fluctuations are Gaussian, the dynamics given by
linear perturbation theory, the tracer bias is scale independent and the
2-point functions are very well measured, the covariance matrix will be
essentially fixed and all models which are close to the data will have very
similar covariance matrices.  Also, for sufficiently constraining data the
assumption of a fixed covariance matrix is relatively good
(see discussion in Appendix \ref{app:cov-p}).
Violations of any of these assumptions (e.g.~not well constrained correlation
functions on large scales or at high redshift, non-linear contributions to the
dynamics, violations of the scale-independent bias assumption or inclusion
of the trispectrum on small scales) can lead to relevent parameter dependence 
of $C$ in the high likelihood regions.
In this regard it is worth noting that it is reasonably straightforward
to include the parameter dependence of $C$ if $C(\vec{p})$ or its inverse
can be evaluated (possibly with regularization, possible simultaneously
\cite{Pou07}) at a few locations and the parameter dependence is smooth
or small.
If the likelihood is insensitive to changes in $C$ with $\vec{p}$ it is
even less sensitive to small interpolation errors when including that
dependence.  The decision on whether to invest the resources to allow
interpolation of the covariance matrix will require a case-by-case analysis.

\subsection{Background}

Both the covariance matrix, $C$, and its inverse, the precision matrix, are
symmetric, positive-definite (SPD) matrices.  Given their importance in so
many branches of engineering and the sciences it is no surprise that there
is a vast literature on the properties of SPD matrices.
The subset of SPD matrices has a rich mathematical structure, being a
manifold with tangent vectors, geodesics and a metric (e.g.~\cite{Hel79}).
In particular, there are several methods for performing interpolation between
such matrices at discretely sampled points in parameter space.  All such
interpolations implicitly depend upon a measure of distance between matrices,
and there are several such distance measures.
We review some of these measures, and the associated interpolations and
averages, in Appendix \ref{app:dist}, and here focus on the results.

Consider linearly interpolating between an SPD matrix $M_0$ at $t=0$ and $M_1$
at $t=1$.  The arithmetic interpolation is trivially $M_t = (1-t)M_0 + t M_1$
and this is also SPD for $0\le t\le 1$ since the SPD matrices form a convex
cone.  The geometric interpolation is almost as simple,
$M_t=M_0(M_0^{-1}M_1)^t$ or the symmetric form
$M_t=M_0^{1/2}(M_0^{-1/2}M_1M_0^{-1/2})^tM_0^{1/2}$,
where the matrix power can be easily done in the diagonal basis.\footnote{Any
SPD matrix can be written $M=UDU^T$ with $D$ diagonal and $D_{ii}>0$, so for
any analytic function $f$ we have $f(M)=Uf(D)U^T$.  We recognize $D$ as the
matrix formed from the eigenvalues and $U$ the matrix formed from the
eigenvectors of $M$.}  In fact precisely this scheme is commonly used in
computer graphics to interpolate perspective changes and rotations
(using quaternions in place of rotation matrices for simplicity).
Unlike arithmetic interpolation, geometric interpolation gives an SPD matrix
for all $t$ and so can also be used for extrapolation although we won't make
use of that feature here.\footnote{In \cite{Mor13}, the authors used a
related property of SPD matrices to allow unconstrained interpolation over
their parameter space without violating the SPD property.}

The generalization to multidimensional interpolation, as described above,
is straightforward.  The arithmetic mean is trivially $m^{-1}\sum_k M_k$.
The geometric mean is not quite so simple, since the matrices may not all
be diagonal in the same basis, however the geometric mean can be defined as
the unique solution, $M$, to $\sum_k {\rm Log}\left(M_k^{-1}M\right)=0$.
%(Ref.~\cite{Moa06} describes an iterative method to solve this equation.)
Interpolation then corresponds to including weights, $u_i$, given by the
areal coordinates of the point in the simplex, i.e.~solving
$M=\sum_k u_k M_k$ or $\sum_k u_k\,{\rm Log}\left(M_k^{-1}M\right)=0$
(see e.g.~\cite{MoaBat06} for further discussion).
We discuss several techniques for solving this second equation in
Appendix \ref{app:geometric}.

\subsection{Diagonal example}

\begin{figure}
\begin{center}
\resizebox{\columnwidth}{!}{\includegraphics{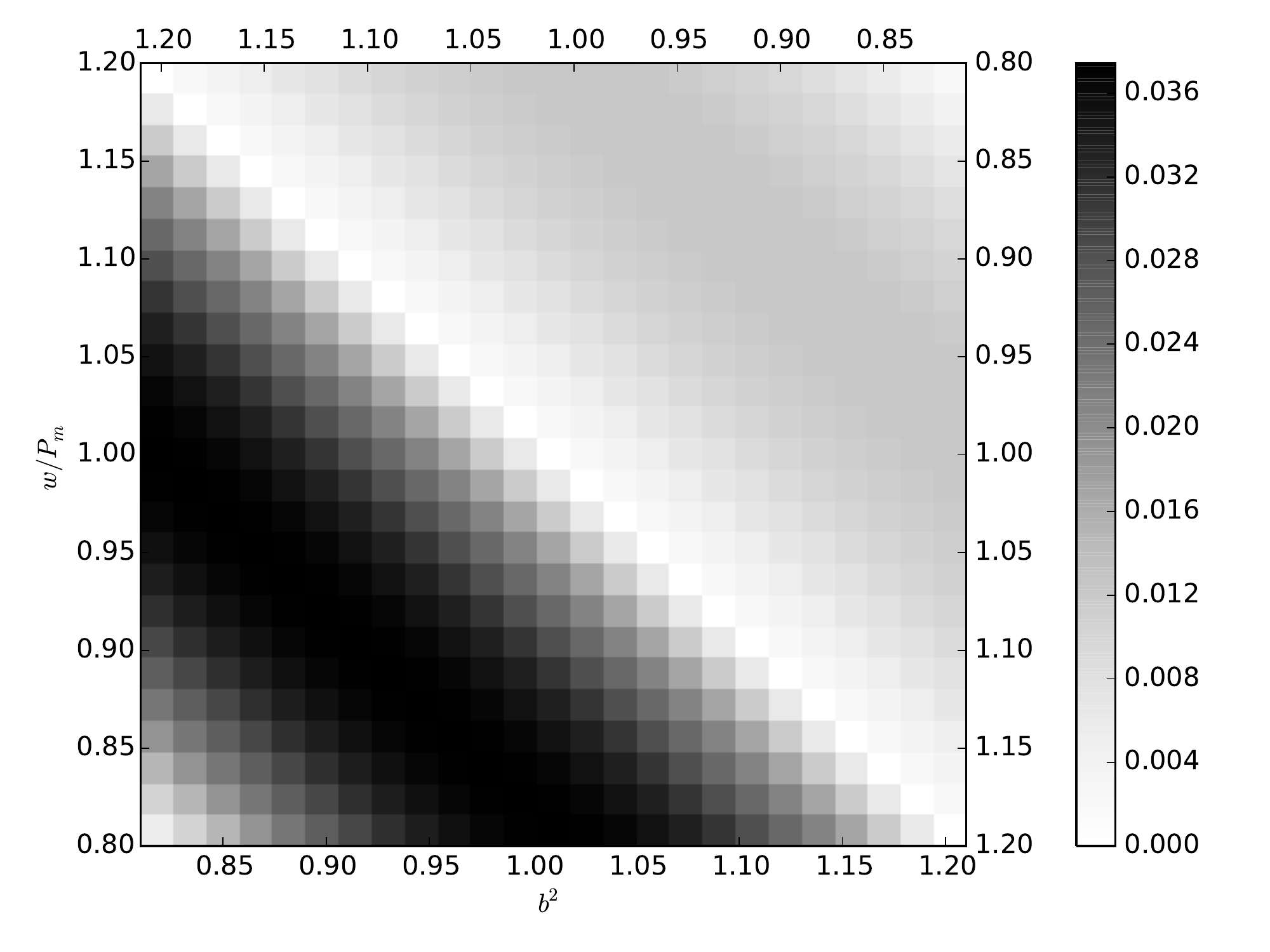}}
\caption{An example of the deviation obtained using arithmetic or geometric
interpolation within a 2D simplex (i.e.~a triangle) for a single element
of the precision matrix corresponding to equation \ref{eqn:lincovP}.
The grey scale shows $\Omega_{\rm est}/\Omega_{\rm true}-1$.
The lower left triangle shows the results for arithmetic interpolation,
while the upper right triangle shows the results for geometric interpolation
(with the upper right vertex being identified with the lower left).
In both cases we interpolate the parameters linearly in $b^2$ and $w$.
We scan the white noise level from below to above the cosmological power since
the interpolations become exact in the limit that either term dominates.
The dark band in the lower triangle indicates the worst-case scenario where
the shot-noise and cosmological power are approximately equal.}
\label{fig:simple_interp}
\end{center}
\end{figure}

Let us illustrate these ideas with some simple examples.
In Gaussian, linear theory, if we work in $k$-space, the matrices are diagonal
and all of these operations become trivial.
A typical Cov in this limit is
\begin{equation}
  {\rm Cov}\left[P_i,P_j\right] = \epsilon_i\left(P_i+w_i\right)^2\delta_{ij}
\label{eqn:lincovP}
\end{equation}
where $P_i$ is the $i$th bandpower, $\epsilon_i=2/N_i$ is a small number when
the number of modes per bin, $N_i=V_{\rm surv}(4\pi k_i^2\Delta k)$, is large
and $w_i=1/\bar{n}$ is the amplitude of the white- or shot-noise component.
In the simplest biasing scheme, and neglecting redshift-space distortions,
$P_i=b^2P_{\rm mat}(k_i)$.

Let us consider the simplest 2D scheme where we vary just $b$ and $w$
and interpolate within the simplex
$(b^2=b_1^2,w=w_1)$, $(b^2=b_2^2,w=w_1)$ and $(b^2=b_1^2,w=w_2)$.
If we were to break equation \ref{eqn:lincovP} into its 3 parts then each
could be interpolated simply, i.e.~both parameters would be `fast'.  We
shall instead consider interpolating the square of the sum as a single block,
treating both $b$ and $w$ as `slow' parameters.
In a modern survey we typically know $b$ and $w$ to order 10-20 per cent,
so we will choose $\pm 20$ per cent as the range over which to vary $b^2$
and $\pm 20$ per cent as the range over which to vary $w$.
For this situation, and for typical values of $b$ and $w$, we find that
both arithmetic and geometric interpolation recover the actual result to
about a per cent.  The deviation is largest furthest from the vertices of the
simplex, and when $b^2P\simeq w$, as expected.
An example is shown in figure \ref{fig:simple_interp}.
We have chosen to interpolate the parameters linearly in $b^2$ and $w$,
though we could also have chosen to interpolate in $\log b^2$ and $\log w$
or any mixture.  The performance is not particularly dependent on this
choice, but for the configuration shown linear-linear interpolation worked
very well.

Although this example is somewhat artificial (we did not split the
covariance matrix into its component parts and we used diagonal matrices)
two characteristics can be seen.  First, we see some preference for the
geometric interpolation over the arithmetic one.  Second, both perform quite
well over the range of parameters we may expect to explore.

\begin{figure}
\begin{center}
\resizebox{\columnwidth}{!}{\includegraphics{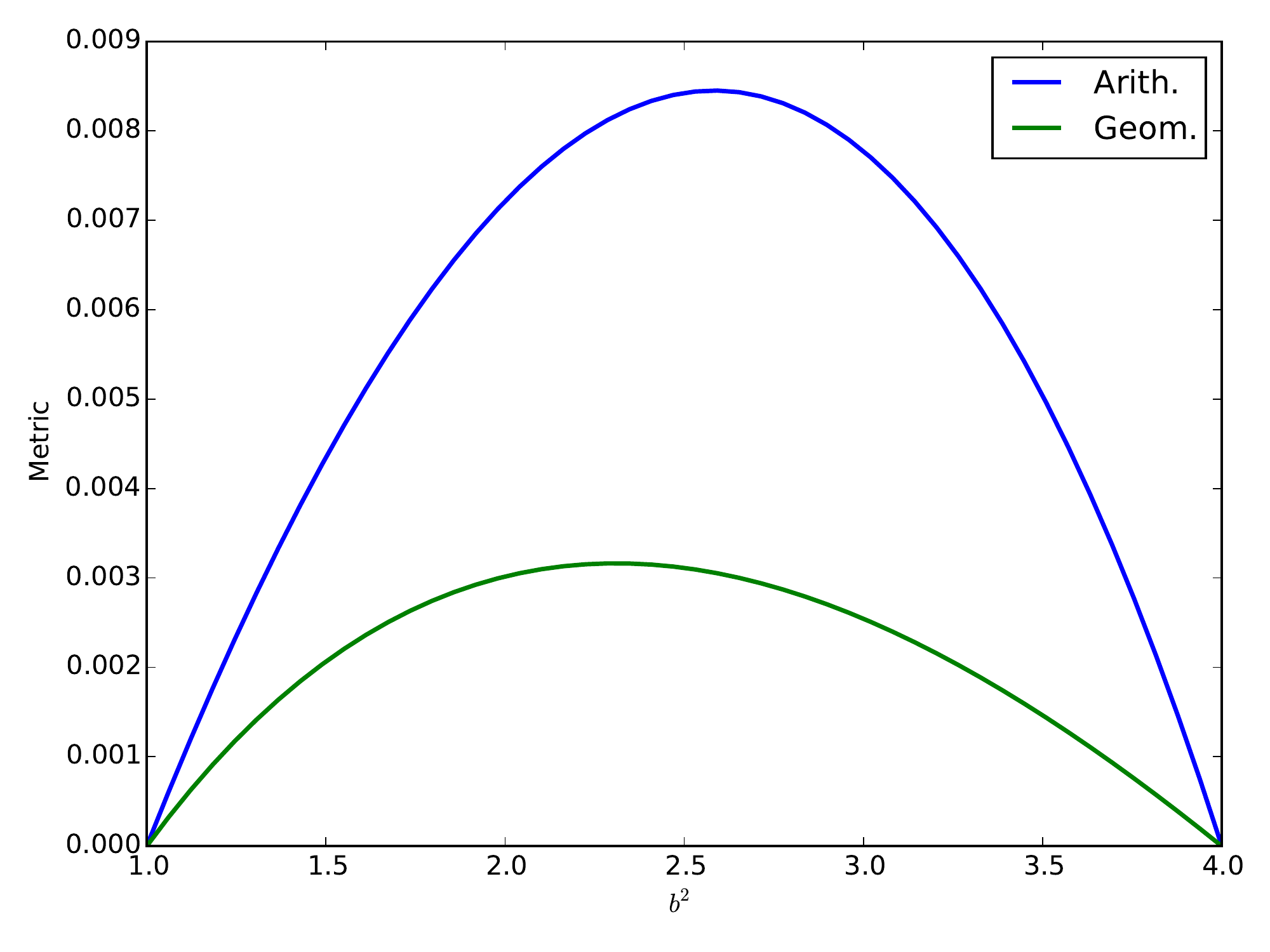}}
\caption{An example of interpolation in one dimension, in this case the
large-scale bias, using arithmetic interpolation and geometric interpolation.
The covariance matrix is $40\times 40$, containing entries for $\xi_0$ and
$\xi_2$ evaluated at 20 points between $s=25\,h^{-1}$Mpc and $125\,h^{-1}$Mpc
using linear theory with $\bar{n}=3\times 10^{-4}\,h^3\,{\rm Mpc}^{-3}$.
The interpolated $n\times n$ precision matrix, $\Omega$, is compared to the
true one using the metric $n^{-1}|| C\ \Delta\Omega ||_F$.
Note the sub-percent deviation even when interpolating over such a broad range
in bias at fixed shot-noise.
For comparison, using a fixed covariance matrix across the range leads to
an order of magnitude larger deviation by the same metric.}
\label{fig:plot_1d}
\end{center}
\end{figure}

\subsection{Matrix example}

\begin{figure}
\begin{center}
\resizebox{\columnwidth}{!}{\includegraphics{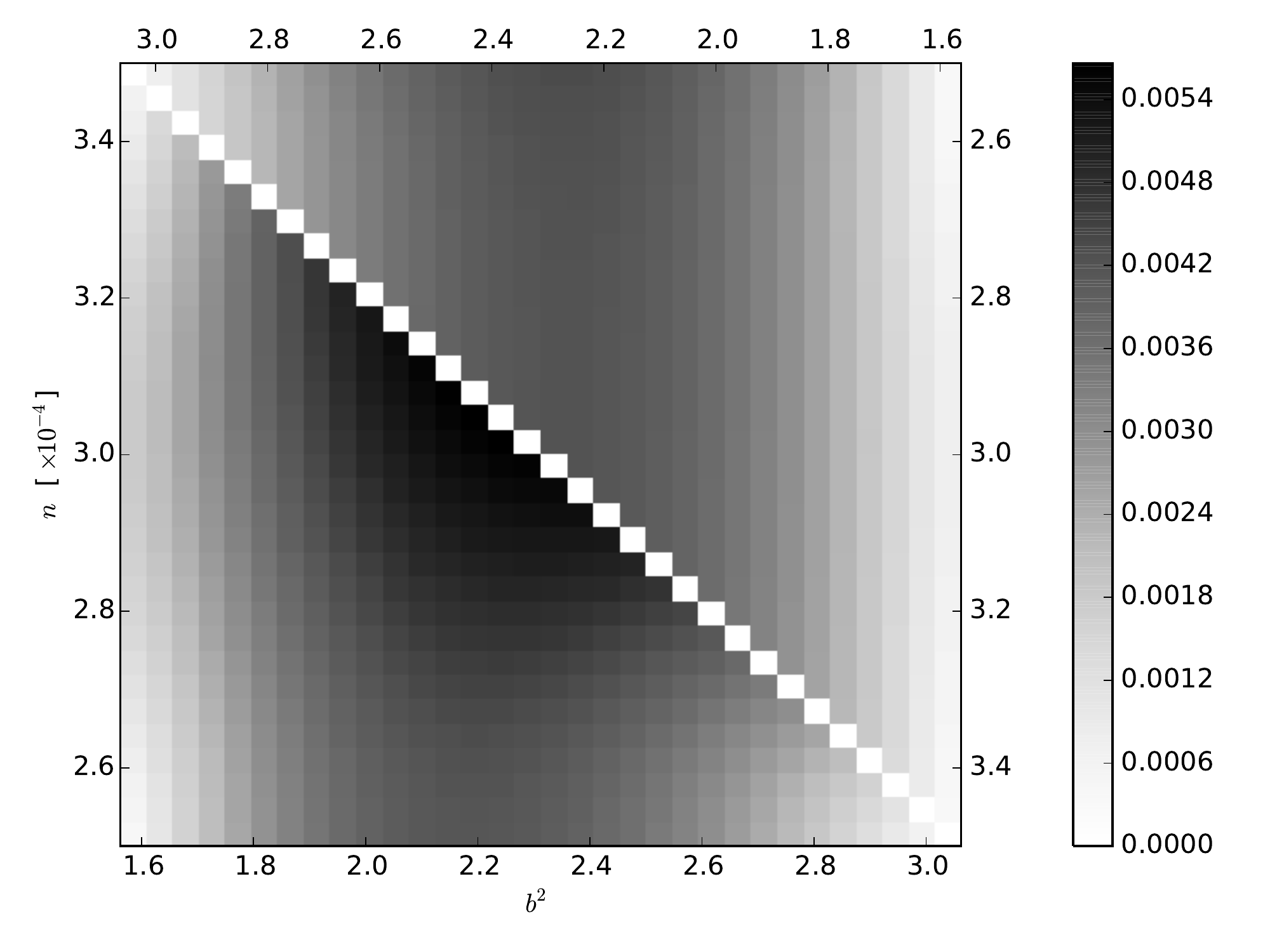}}
\caption{An example of interpolation in two dimensions (the large-scale bias
and the number density).
The covariance matrix is $40\times 40$, containing entries for $\xi_0$ and
$\xi_2$ evaluated at 20 points between $s=25\,h^{-1}$Mpc and $125\,h^{-1}$Mpc
computed using linear theory.
The greyscale compares the interpolated $n\times n$ precision matrix,
$\Omega$, to the true one using the metric $n^{-1} || C\ \Delta\Omega ||_F$.
The lower right half gives the results for arithmetic interpolation while the
upper right gives the results for geometric interpolation (with the upper
right vertex identified with the lower left).  See text for further discussion.}
\label{fig:plot_2d}
\end{center}
\end{figure}

This exercise was artificial in dealing with diagonal matrices.  As a
next step let us consider interpolation of non-sparse matrices.
For this example we use the covariance matrix for the multipoles of the
correlation function, computed in linear theory assuming
Gaussian statistics and normalized by the survey volume, $V$.
Motivated by the example of fitting baryon acoustic oscillations and
redshift-space distortions we compute the covariance of $\xi_\ell(s)$
in linear theory in 20 bins of $s$ running from $25\,h^{-1}$Mpc to
$125\,h^{-1}$Mpc for $\ell=0$ and $\ell=2$.
This results in a $40\times 40$ covariance matrix with neighboring bins
quite highly correlated.

We interpolate the full matrix (i.e.~not dividing it into shot-noise and
sample variance terms or pre-whitening it in any way) in $b^2$ holding the
shot-noise fixed and compare the interpolated value to the exact calculation.
The result is shown in figure \ref{fig:plot_1d} where we see a clear
preference for geometric interpolation over arithmetic interpolation but
excellent performance from both.
The metric used to compare the interpolated, $n\times n$ precision matrix
to that evaluated explicitly is
$n^{-1}|| C\,\Delta\Omega ||_F$, where $||\cdot||_F$
indicates the Frobenius norm (see Appendix \ref{app:dist}).
This is essentially the rms (averaged over elements) deviation from the
identity of $\Omega_{\rm est}$ times $C_{\rm true}$.
We see that interpolation induces sub-percent rms deviations even over such
a broad range.
For comparison, using a fixed covariance matrix across the range leads to
an order of magnitude larger difference at the endpoints by the same metric.
If we interpolate the correlation matrix (i.e.~the covariance matrix normalized
to unity along the diagonals) then the deviation is approximately halved, but of
course we must then additionally interpolate the scaling of the diagonal
entries.

Next we consider interpolating in $b^2$ and $\bar{n}$.
In this case we interpolate over the range $1.25<b<1.75$
and $2.5\times 10^{-4}<\bar{n}<3.5\times 10^{-4}\,h^3{\rm Mpc}^{-3}$.
The results are shown in figure \ref{fig:plot_2d} with the same metric as
in figure \ref{fig:plot_1d}.  In this situation the preference for geometric
interpolation over arithmetic interpolation is less clear, though we do find
that the geometric interpolation does tend to perform better with the chosen
metric.
Again the absolute performance of both the arithmetic and geometric schemes
is very good.
If we had used a fixed covariance matrix (e.g.~the one at the lower left
corner) the metric at the worst point is 10 times larger than the worst-case
shown in figure \ref{fig:plot_2d}.

An alternative metric for evaluating how well we have estimated $\Omega$ is
to compute tr$[C_{\rm true}\Omega_{\rm est}]-n$ where $C$ and $\Omega$ are
$n\times n$ matrices.  If the data are drawn from a Gaussian distribution,
centered on the theory and with covariance $C_{\rm true}$, and if $\chi^2$
is computed using $\Omega_{\rm est}$ for the precision matrix then the
average value of $\chi^2$ is tr$[C_{\rm true}\Omega_{\rm est}]$.
For the case considered in figure \ref{fig:plot_2d},
tr$[C_{\rm true}\Omega_{\rm est}]-n$ is never larger than about $0.03\,n$
and usually considerably less.
This can be compared to the $\sqrt{2n}$ change in $\chi^2$ required for a
$1\,\sigma$ change in the likelihood.
On average, then, the error made by using the interpolated $\Omega$ in
place of the `true' $\Omega$ in this case is negligible compared to the
statistical uncertainties in the measurement and probably small compared
to the errors in determining the $C_k$ by Monte Carlo at each of the
control points.
We have investigated a number of other metrics for comparing matrices, and
find a similar story by each measure.

\subsection{More dimensions and simplices}

As a final example we consider interpolating the covariance matrix for
the multipoles of the correlation function, as above, in a 4-parameter
space and over multiple simplices.  We allow $b$ and $\bar{n}$ to vary,
as above, but now additionally include variations in the linear growth rate,
$f$, and a dilation parameter $\alpha$.  The large-scale bias and shot-noise
affect the amplitudes of the two contributions to the covariance, the growth
rate alters the the quadrupole to monopole ratio and the dilation parameter
shifts the entire correlation function in scale.  Thus this parameter set
allows considerable freedom in the structure of the matrices we are
interpolating over.

We take the range of the parameters to be
$1.25<b<1.75$,
$2.5<\bar{n}<3.5\times 10^{-4}\,h^3\,{\rm Mpc}^{-3}$,
$0.74<f<0.76$ and
$0.98<\alpha<1.02$.
To start, we evaluate the covariance matrices at points distributed throughout
the volume (see below).
Then we throw points at random within the space.  For each point we find its
enclosing simplex, compute the areal coordinates within that simplex and
perform geometric interpolation from the 5 vertices of the simplex.
Using the same metric as before, we find that the performance is excellent.

We have investigated several distributions of points, finding that they all
perform quite well.  First we placed the points at the vertices of a hypercube
bounding the parameter set (i.e.~picking either the upper or lower limits,
above, for each parameter), plus an additional point at the center for a total
of 17 points.
The rms deviation of $C_{\rm true}\Omega_{\rm est}$ from the identity matrix
for randomly thrown points was sub-percent.
We also investigated using orthogonal arrays of points with 8 or 9 vertices
(specifically OA(8,4,2,3) and OA(9,4,3,2) purely as illustrative examples).
Within the region bounded by the points the rms deviation as again below a
percent.
Finally we investigated a 9 point configuration with a central point and
8 points where each parameter was individually varied to the edges of its
range while the others were held fixed at the central point.  Again we
found that the interpolation performed very well for the enclosed points.
For every case we investigated with approximately 10 points, sub-percent
rms deviations were found throughout the region enclosed by the points.

Our conclusion is thus that low order interpolation from an unstructured
mesh, while it may not be optimal, is very likely sufficient for upcoming
surveys focused on BAO and RSD analyses of the power spectrum and
correlation function.  Using an interpolated precision matrix can reduce
the error in the likelihood, compared to holding the precision matrix fixed,
by a non-trivial factor and even a relatively coarse set of points
provides an accurate interpolation.
We have not considered the interesting problem of optimizing the position of
the points in the mesh.  Any such solution is likely to be problem specific,
and to depend on the nature of the matrices being interpolated.  Instead we
have shown that the performance of our simple interpolation is not strongly
dependent on the choice of mesh points.
Finally, it is very conceivable that a more sophisticated interpolation
approach could be devised and that it would save computational resources.
The geometric interpolation method we have advocated has a simple
generalization to techniques such a kriging, inverse distance weighting
or kernel estimation.
We leave such investigations to future work.

\section{Conclusions}
\label{sec:conclusions}

The study of large-scale structure has become a powerful means of learning
about cosmology and fundamental physics.  Analyses of large-scale structure
data sets have become increasingly complex and sophisticated as the data
themselves have grown in size and constraining power.
In this paper we have focused on the comparison of theoretical models to
the most common summary statistics derived from such data: the redshift-space
power spectrum and correlation function.
Under the assumption of a Gaussian likelihood function the computation of
statistical confidence levels or goodness-of-fit of models requires
knowledge of the theory, the data and the covariance matrix.  In principle
all three can depend on the parameters being tested though the full
dependence is not usually included.
We have shown that a very simple interpolation scheme from an unstructured
mesh allows for an efficient way to include this parameter dependence
self-consistently in the analysis at modest computational expense.

We have advocated multi-linear interpolation, under the assumption that the
range of parameters being explored is relatively small and the changes
smooth.  For vector-valued quantities, such as the data vector of $P(k_i)$
or $\xi_\ell(s_i)$, the interpolation is straightforward.
For matrix-valued quantities the best interpolation scheme is not as
obvious.  We have compared two, arithmetic interpolation and geometric
interpolation, and found them to perform well though the geometric scheme
(which uses the geometric structure of the group) can have up to half of
the deviation of the simpler arithmetic scheme.
As an illustrative example we showed that, for reasonable variations in
shot-noise and large-scale bias, sub-percent deviations from interpolation
could be achieved in predicting the linear theory covariance matrix at any
point interior to the simplex.
The interpolation introduces errors significantly smaller than the
statistical uncertainty.

This work assumes that the data and covariance matrix have been computed
at the simplex points.  We leave to future work an investigation of the
optimal mesh placement and efficient ways to compute the covariance matrices
by conditioning the sample covariance derived from Monte Carlo simulation.

\noindent \emph{Acknowledgements}:
NP is supported in part by a DOE Early Career grant DE-SC0008080.
MW would like to thank the Royal Observatory of Edinburgh and the
Higgs Centre for their hospitality while much of this work was completed.
We would like to thank Joanne Cohn and Avery Meiksin for helpful conversations
on group theory and numerical methods.

\appendix

\section{Neglecting the parameter dependence of the covariance}
\label{app:cov-p}

For large data sets, with well constrained parameters, it is often
possible to neglect the parameter dependence of the covariance matrix
compared to that of the mean, or theoretical prediction.
This is obviously the case if the covariance matrix is independent of
the parameters (e.g.~the shot-noise limit for galaxy surveys) but can be
true more generally.
In this situation performing a simple $\chi^2$ analysis is enough.
In this appendix we quantify how neglecting the parameter dependence of the
covariance matrix affects the likelihood in some simple cases, always
assuming that the `fixed' covariance matrix is chosen to be close to that
of the best fit model.  For a discussion of how an incorrect covariance
matrix affects the likelihood, see e.g.~\cite{Tay13,Mor13}.

As in the main text, let us assume a Gaussian form for the likelihood of
the theory, $\mu$, given data, $d$, with covariance $C$:
\begin{equation}
  -2\ln\mathcal{L} = (d-\mu)^T C^{-1} (d-\mu) + {\rm tr}\ln C + \cdots
\end{equation}
then the parameter dependence can enter in the mean, $\mu(p)$, or variance,
$C(p)$.  Obviously we cannot make a generic statement about which is the
dominant dependence, since e.g.~a priori we could have $\mu$ or $C$ be
$p$-independent.  For the case of galaxy surveys however we can make
considerable progress
(see \cite{Teg97} for very related discussion and
 \cite{LabStaLac12} for an investigation for the case of BAO and
 \cite{Eif09} for an investigation for the case of cosmic shear).

Let us consider how the likelihood function falls from its peak as we vary
the parameters away from the best-fit point.  For analytic simplicity and
in order to bring out the key points we shall work in linear theory and
use a $k$-space basis.  Without loss of generality we can take our `parameters'
to be bandpowers in a $k$-bin, $p_i$, with the measurements of the power
spectrum in the bin being $P_i=P(k_i)$.
With these assumptions $\mu=p_i$ while
Cov$[P_i,P_j]=\epsilon_i(p_i+w_i)^2\delta_{ij}$ where $\epsilon_i=2/N_i$ is a
small number when the number of independent modes per bin, $N_i$, is large
and we have written the white- or shot-noise component as $w_i$.
Our covariance matrix is most parameter dependent when $w_i\equiv 0$, or we
are in the `sample variance limit'.  We shall therefore make this assumption.
Note that the covariance is diagonal in $k$-space in linear theory, so the
log-likelihood becomes
\begin{equation}
  -2\ln\mathcal{L} = \sum_i  \ln|C_{ii}| + \frac{(P_i-p_i)^2}{C_{ii}} + \cdots
  = \sum_i \ln p_i^2 + \frac{(P_i-p_i)^2}{\epsilon_i p_i^2} + \cdots
\end{equation}
where $\cdots$ represents terms independent of $p_i$.
Note how the first term (from the variation of the covariance matrix) is
independent of $\epsilon_i$ while the second becomes increasingly more
important as $\epsilon_i\to 0$.
For large amounts of data, small $\epsilon_i$, the parameter dependence of the
covariance term only becomes important once we are far from the peak of the
likelihood function, i.e.~for models which are disfavored by the data.
To see this, note that for a model to be allowed $p_i$ must be close to $P_i$,
i.e.~within $\sqrt{\epsilon}$.  For such variations, we can compute the
difference between allowing the covariance to vary and holding it fixed
at the peak of the likelihood.  This difference scales as $\sqrt{\epsilon}$.
The parameter dependence of the covariance therefore only becomes important
for models disfavored by the data.

An alternative approach \cite{Teg97} is to look at the Fisher matrix, i.e.~the
expectation value of the Hessian of the log-likelihood.  For a Gaussian this
becomes
\begin{equation}
  F_{\alpha\beta} \propto {\rm tr}\left[C^{-1}C_{,\alpha}C^{-1}C_{,\beta}\right]
  + \mu_{,\alpha} C^{-1} \mu_{,\beta}
\end{equation}
where $,\alpha$ denotes a derivative with respect to parameter $p_\alpha$ and
we have chosen Greek indices to label the parameters which can now be arbitrary.
The Fisher matrix represents the curvature of the likelihood around the peak,
assuming `typical' data.
Once again, note that the first term (arising from the parameter-dependence
of the covariance) is independent of the overall normalization of $C$ while
the second term scales as the inverse of the normalization.
It is easy to show that for bandpowers as parameters and the same assumptions
as above the first term is $\mathcal{O}(\sum_i p_i^{-2})$ while the second is
$N_i$ times larger and so dominates in the large-volume, high-precision limit.

\section{Distance measures for SPD matrices}
\label{app:dist}

In this appendix we quickly review some common distance measures on the space
of symmetric, positive-definite (SPD) matrices.

The most basic distance measure is the Frobenius norm (sometimes called the
Hilbert-Schmidt norm).
The Frobenius norm is the one induced by the inner product on matrices
thought of as a vector space:
$\langle \mathbf{A},\mathbf{B}\rangle={\rm tr}(\mathbf{AB}^T)$.
If we work with symmetric matrices,
$|| \mathbf{A} ||=\sqrt{{\rm tr}(\mathbf{A}^2)}$.

We can also use the Kullback-Liebler divergence (or information gain).
The KL divergence\footnote{A divergence is a generalization of a metric
which need not be symmetric or satisfy the triangle inequality.
Formally a divergence on a space $X$ is a non-negative function on the
Cartesian product space, $X\times X$, which is zero only on the diagonal.}
between two zero-mean Gaussians with covariances $C_a$ and $C_b$ is
\begin{equation}
  D_{KL}(C_a||C_b) = \frac{1}{2}\left[ {\rm tr}\left(C_a C_b^{-1}\right)
   -\log\det\left(C_a C_b^{-1}\right) - N \right]
\end{equation}
with $C_{a}$ and $C_{b}$ $N\times N$ matrices and $D\ge 0$ with equality
only if $C_a=C_b$.
A distance measure can be constructed from $D_{KL}$ by symmetrizing the
arguments.  If $\lambda_i$ are the eigenvalues of $C_a^{-1}C_b$ then
\begin{equation}
  D_{KL}^{\rm sym}(C_a,C_b) = \frac{1}{2}\sum_{i=1}^N
  \left(\sqrt{\lambda_i}-\frac{1}{\sqrt{\lambda_i}}\right)^2
\end{equation}
and the square root of $D_{KL}^{\rm sym}$ is a distance measure which is
invariant under congruent transformations and inversion.

Finally we can consider the space of positive-definite symmetric matrices as
a manifold, with a Riemannian metric defined by the Frobenius dot product on
the tangent space at any point.  The geodesic distance between matrices $C_a$
and $C_b$ then becomes \citep[e.g.][]{MoaBat06}
\begin{equation}
  D_R(C_a,C_b) = || {\rm Log}(C_a^{-1}C_b) ||_F
               = \sqrt{\sum_{i=1}^N \log^2\lambda_i}
\end{equation}
where the $\lambda_i$ in the last equality are the eigenvalues of the matrix
$C_a^{-1}C_b$.
Note that for any positive-definite symmetric matrix $M$ we can decompose it
into a diagonal and orthogonal matrix as $M=UDU^T$ with $D$ containing the
eigenvalues and $U$ the eigenvectors of $M$.  In this basis taking Log$M$
simply becomes taking the log of the diagonal entries,
i.e.~${\rm Log}M=U\ln(D)U^T$.

Each of these different distance measures naturally lead to a different
interpolation scheme.
Consider for example taking the `average'.  Under the Frobenius norm the
arithmetic average, $\bar{C}$, of the matrices $C_k$ minimizes the squared
distance: $\sum_k\left|\bar{C}-C_k\right|^2$.
The geodesic distance leads to geometric\footnote{It is easy to see why this
is heuristically.  For a Lie manifold the tangent vectors are isomorphic to
the group generators.  Since group generators and elements are related by
exponentiation, the vectors are logs of matrices and lengths along parallel
transported vectors are `logarithmic' which converts an arithmetic mean into
a geometric one.  In fact the geodesic running from the origin to a group
element $G$ is simply $\exp[t\,{\rm Log}G]$ for $t\in [0,1]$.  The geodesic
running from $P$ to $Q$ is $P^{1/2}(P^{-1/2}QP^{-1/2})^tP^{1/2}$.  See
\cite{Smi05} for a recent discussion of covariance matrices within the
context of manifolds and a list of references.} means of
matrices as the appropriate average, while the KL-based distance gives the
geometric mean of the arithmetic and harmonic means \cite{Moa06}.
Similar situations arise when performing interpolation, as described in the
main text.

\section{Solving the geometric interpolation equation}
\label{app:geometric}

Suppose we are given the values of the covariance matrix at the
$m=n_p+1$ corners of a simplex, $C_k$, and the areal coordinates
$u_k$ corresponding to the (interior) point for which we want the
interpolated value of the precision matrix, $\Omega$.
In order to interpolate geometrically we need to find the matrix
$\Omega$ which solves
\begin{equation}
  G(\Omega) \equiv \sum_{k=1}^m u_k {\rm Log}\left(C_k\Omega\right) = 0
  \quad .
\label{eqn:geom_eqn}
\end{equation}
One method would be to use Newton's algorithm for matrix valued functions
of matrices.  This requires the (Fr\'{e}chet) derivative of $G$.
Fast numerical algorithms for computing both the
logarithm\footnote{Interestingly, the algorithm is based on the one used by
Briggs in his heroic ``Arithmetica Logarithmica'', published in 1624.}
and its derivative exist \cite{Frechet}, but are complex.
A more standard approach is to use a variant of Broyden's method \cite{Bro69},
or other quasi-Newton methods, which do not require the explicit evaluation
of a derivative.  By stacking the columns of our matrices atop each other we
can turn $n\times n$  matrices into $n^2$-dimensional vectors (this is the
``vec'' operation, equivalent to replacing pairs of indices, $ij$, by a
super-index, $I$) and the problem reduces to a standard one of simultaneously
solving multiple non-linear equations.
Various efficient and general algorithms for this problem exist, however
convergence can be difficult to achieve when the number of terms in the sum
and the dimension of the matrices becomes large.

The method we have used is a fixed point algorithm due to ref.~\cite{Moa06}.
If we rotate our bases by $\Omega_1^{-1/2}$ to set $\Omega_1=1$ and left and
right multiply by $\Omega^{1/2}$ and $\Omega^{-1/2}$ we can write equation
\ref{eqn:geom_eqn} as
\begin{equation}
  {\rm Log}\hat{\Omega} = -\sum_{k=2}^m u_k'\ {\rm Log}
  \left(\hat{\Omega}^{1/2}\hat{C}_k\hat{\Omega}^{1/2}\right)
\end{equation}
where hats denote the matrices in rotated coordinates and the $u_k'$ are
$u_k$ divided by $1-\sum_{k=2}^m u_k$.
Ref.~\cite{Moa06} shows that this can be solved for $S={\rm Log}\hat{\Omega}$
using a fixed point iteration
\begin{equation}
  S^{\ell+1}=\alpha S^\ell + (\alpha-1)
  \sum_{k=2}^m u_k'\ {\rm Log}
  \left( \exp(S^\ell/2)\hat{C}_k\exp(S^\ell/2)\right)
\end{equation}
starting at $S^{0}=\sum_k u_k'\ {\rm Log}\hat{\Omega}_k$.
This converges for $\alpha=(m+a)/(m+a+1)$ for any integer $a>0$ and we have
found that convergence is typically very rapid.
In fact for the situations we have explored the starting guess is already
a good approximation to the geometric mean.
When checking for convergence it is easier to use the symmetric form
\begin{equation}
  \Omega^{1/2}G(\Omega)\Omega^{-1/2} = 0 =
  \sum_{k=1}^m u_k {\rm Log}\left(\Omega^{1/2}C_k\Omega^{1/2}\right)
\end{equation}
for which the arguments of the Log are explicitly SPD, which allows for
quick computation.

%\section{Models for covariance}
%
%[{\bf Remove this?}]
%
%Models for covariance matrices:
%\begin{enumerate}
%\item Linear theory.
%\item Smooth Gaussian model \cite{Xu12}
%\item Analytic model \cite{deP12}
%\item Direct Monte-Carlo simulation
%\item The above plus some regularization scheme.
%\end{enumerate}
%
%Here we can also ask how good is the assumption of varying the diagonal
%errors while keeping the correlation matrix unchanged -- this is often
%done but I'm not sure what the justification is.

%\section{Gibbs}
%
%Want to think a bit about whether we can use the interpolation/emulator
%idea with our old Gibbs sampling idea from May 2014.

\bibliographystyle{JHEP}
\bibliography{ms}
\end{document}